\documentclass[11pt]{article}

\usepackage[T1]{fontenc}
\usepackage{tgpagella}

\usepackage[top=2cm, bottom=2cm, left=2cm, right=2cm]{geometry}

\usepackage{amsmath}
\usepackage{graphicx}
\usepackage{paralist}
\usepackage{indentfirst}

\usepackage[sort&compress,numbers]{natbib}

\usepackage{algorithm}
\usepackage[small,bf]{caption}

\usepackage{flushend}
\usepackage{color}
\usepackage[hyphenbreaks]{breakurl}
\definecolor{darkred}{RGB}{153,0,0}
\definecolor{darkblue}{RGB}{0,0,99}
\usepackage[colorlinks=true, linkcolor = darkred,   citecolor = darkred, urlcolor = darkred]{hyperref}

\usepackage{hyperref}
\usepackage{color, colortbl}
\makeatletter
\def\BState{\State\hskip-\ALG@thistlm}
\makeatother

\definecolor{Gray}{gray}{0.9}

\renewenvironment{thebibliography}[1]{
  \begin{oldthebibliography}{#1}
    \setlength{\itemsep}{0.2em}
    \setlength{\parskip}{0.0em}
}
{
  \end{oldthebibliography}
}

\newcommand{\descr}[1]{\medskip \noindent \textbf{#1}}
\newcommand{\descrit}[1]{\smallskip \noindent \textit{#1}}

\usepackage{url}
\usepackage[hyphenbreaks]{breakurl}

\makeatletter
\def\url@leostyle{%
  \@ifundefined{selectfont}{\def\UrlFont{}}%
  {\def\UrlFont{}}%
}
\makeatother  
\begin{document}

\title{\bf Synthetic Data: Methods, Use Cases, and Risks}

\author{Emiliano De Cristofaro, University of California, Riverside\\[0.25ex]
emilianodc@cs.ucr.edu}

\date{}

\maketitle

\begin{abstract}
Sharing data can often enable compelling applications and analytics.
However, more often than not, valuable datasets contain information of sensitive nature, and thus sharing them can endanger the privacy of users and organizations.
A possible alternative gaining momentum in both the research community and industry is to share {\em synthetic data} instead.
The idea is to release artificially generated datasets that resemble the actual data -- more precisely, having similar statistical properties.

In this article, we provide a gentle introduction to synthetic data and discuss its use cases, the privacy challenges that are still unaddressed, and its inherent limitations as an effective privacy-enhancing technology.
\end{abstract}

\section*{How To Safely Release Data?}
We begin by reviewing the traditional ``alternative'' steps and technologies used by practitioners to share data while attempting to reduce information leakage.

\descrit{Anonymization:} Theoretically, one could remove personally identifiable information before sharing it.
However, in practice, anonymization fails to provide realistic privacy guarantees because a malevolent actor often has auxiliary information that allows them to re-identify anonymized data.
For example, when Netflix de-identified movie rankings (as part of a challenge seeking better recommendation systems), Arvind Narayanan and Vitaly Shmatikov~\cite{narayanan2009anonymizing} were able to de-anonymize a significant portion by cross-referencing them with public information on IMDb.

\descrit{Aggregation:} Another approach is to share aggregate statistics about a dataset.
For example, telecommunication providers can provide statistics about how many people are in some specific locations at a given time — e.g., to assess footfall and decide where one should open a new store.
However, this is often ineffective too~\cite{apostolos,pyrgelis2017knock}, as the aggregates can still help an adversary learn something about specific individuals.

\descrit{Differentially Private Statistics:} More promising attempts come from providing access to statistics obtained from the data while adding noise to the queries’ response, guaranteeing %
differential privacy (DP)~\cite{nissim2017differential}.
DP provides mathematical guarantees against what an adversary can infer from learning the result of some algorithm; i.e., it guarantees that an individual will be exposed to the same privacy risk whether or not her data is included. %
DP is generally achieved by adding noise at various steps of training.

However, this approach generally lowers the dataset's utility, especially on high-dimensional data.
Additionally, allowing unlimited non-trivial queries on a dataset can reveal the whole dataset, so this approach needs to keep track of the privacy budget over time.

\section*{Types of Synthetic Data}
There are different approaches to generating synthetic data.
Derek Snow of the Alan Turing Institute lists %
three main methods~\cite{snow}:

\begin{enumerate}
\item {\em Hand-engineered methods} identify an underlying distribution from real data using expert opinion and seek to imitate it.

\item {\em Agent-based models} establish known agents and allow them to interact according to prescribed rules hoping that this interaction would ultimately amount to distribution profiles that look similar to the original dataset.

\item {\em Generative machine models} learn how a dataset is generated using a probabilistic model and create synthetic data by sampling from the learned distribution.
\end{enumerate}

In the rest of this article, we will focus on {\bf\em generative models}, as they are generally considered state-of-the-art.

\section*{Background: Generative vs.~Discriminative Models}

A good way to understand how generative models work is to look at how they differ from discriminative models.
Consider, for instance, the task of identifying which paintings are by Vincent Van Gogh.
First, we label a dataset of artworks we know whether or not were painted by Van Gogh.
Then, we train a {\em discriminative} model to learn that specific characteristics (e.g., colors, shapes, or textures) are typical of Van Gogh.
We can now use that model to predict whether Van Gogh authored any painting.\\

\begin{figure}[h]
\centering
\includegraphics[width=0.85\columnwidth]{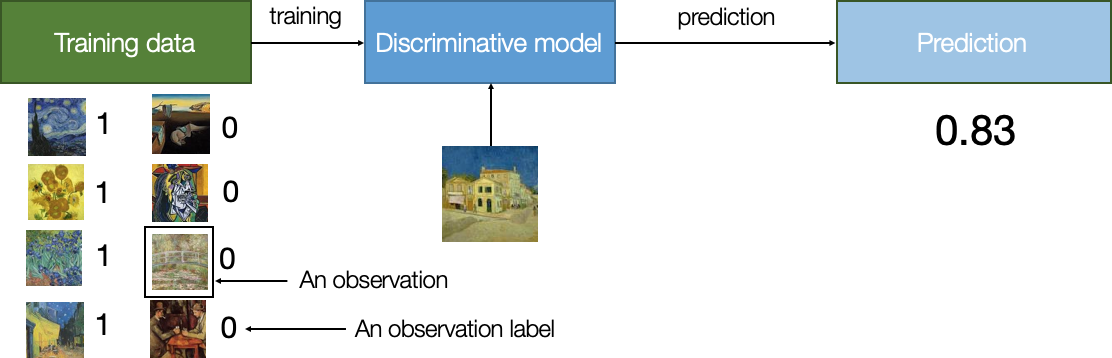}
\caption{{\em Discriminative} Machine Learning Models. (Source: ``\href{https://learning.oreilly.com/library/view/generative-deep-learning/9781492041931/}{Generative Deep Learning,}'' (CC BY 4.0).}
\end{figure}

Whereas generative models allow the generation of a new image of a horse that does not actually exist but looks real.
More precisely, we can train a {\em generative} model to learn {\em what horses look like}; to do so, we need a dataset with many examples (observations) of horses.\\

\begin{figure}[h]
\centering
\includegraphics[width=0.8\columnwidth]{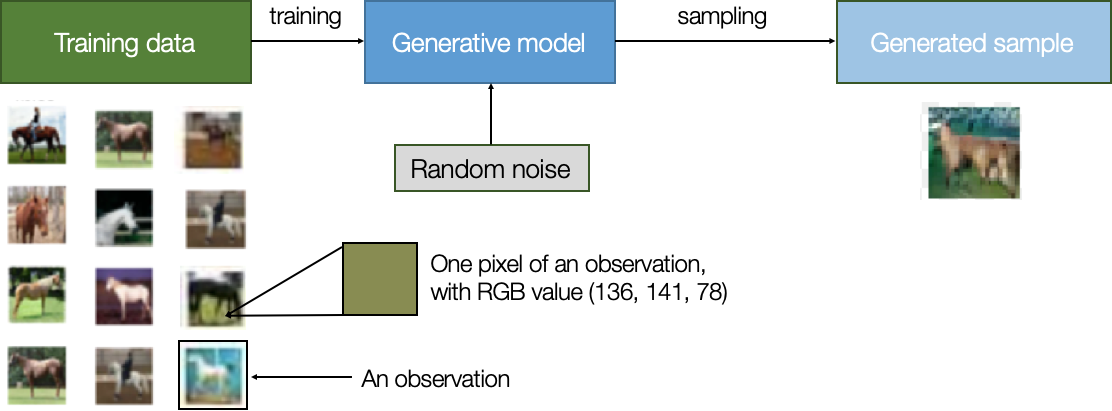}
\caption{{\em Generative} Machine Learning Models. (Source: ``\href{https://learning.oreilly.com/library/view/generative-deep-learning/9781492041931/}{Generative Deep Learning,}'' CC BY 4.0).}
\end{figure}

Each observation has many characteristics (or {\em features}), e.g., each pixel value.
The goal is to build a model that can generate new sets of features that look like they have been created using the same rules as the original data.

\section*{Algorithms}
Generative models used to produce synthetic data may use a number of architectures.
For instance, Generative Adversarial Networks, or GANs~\cite{goodfellow2014generative}, are commonly used to generate artificial images, videos, etc.
The intuition is to pit two neural networks against each other: a generator produces real-looking images while the discriminator tries to distinguish between real and fake images.
The process ends when the discriminator can no longer discern.\\

\begin{figure}[h]
\centering
\includegraphics[width=0.65\columnwidth]{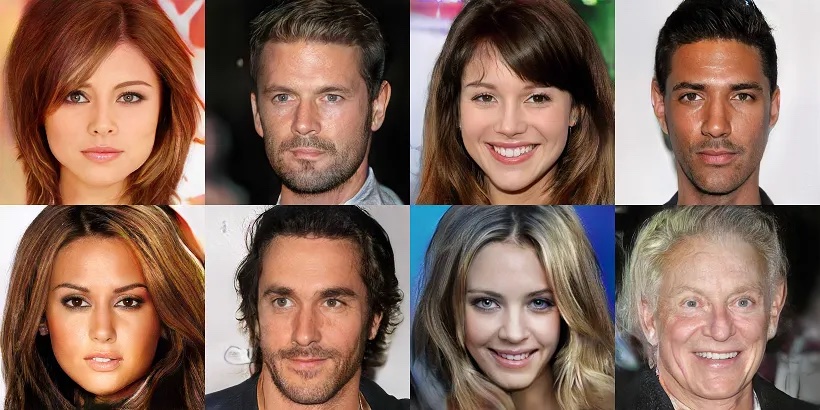}
\caption{{\em Generative} GAN-generated, artificial images. (Source: \cite{nvidia})}
\end{figure}

Besides GANs, there are several other architectures, often based on generative models, used to produce synthetic data.
For instance, Variational Autoencoders try to compress the data to a lower dimensional space and then reconstruct it back to the original.
More methods include Restricted Boltzmann Machines, Bayesian networks, %
Markov chain Monte Carlo methods, etc.

\section*{What Can Synthetic Data Be Used For?}
Nowadays, there are a number of companies %
that market synthetic data technologies, e.g., by \href{https://datagen.tech}{Datagen}, \href{https://mostly.ai}{Mostly.ai}, \href{https://hazy.com}{Hazy}, \href{https://gretel.ai}{Gretel.ai}, and \href{https://aindo.com}{Aindo}.
They mention several use cases, including:

\begin{enumerate}
\item {\em Training Machine Learning Models:} synthetic data can be used to augment real data, upsample/rebalance under-represented classes, or make models more robust to special events, e.g., in the context of fraud detection~\cite{mostly}, healthcare~\cite{tucker2020generating}, etc. 

\item {\em Product and Software Testing:} generating synthetic data can be easier than obtaining and using real rule-based test data to assist %
with various testing tasks.
For example, companies are often unable to legally use production data for testing purposes.

\item {\em Governance:} synthetic data can help remove biases, stress-test models, and increase explainability.

\item {\em Privacy:} synthetic data can mitigate privacy concerns when sharing or using data across and within organizations.
Datasets are considered ``anonymous,'' ``safe,'' or void of personally identifiable information.
This allows data scientists to comply with data protection regulations like \href{https://en.wikipedia.org/wiki/Health_Insurance_Portability_and_Accountability_Act}{HIPAA} (in the US), \href{https://en.wikipedia.org/wiki/General_Data_Protection_Regulation}{GDPR} (in the EU), or \href{https://en.wikipedia.org/wiki/California_Consumer_Privacy_Act}{CCPA} (in California), etc.
\end{enumerate}

Overall, over the past few years, there have been several initiatives and efforts both in industry and government.
For example, the UK's National Health Service piloted a project to \href{https://data.england.nhs.uk/dataset/a-e-synthetic-data}{release synthetic data} from ``A\&E'' (i.e., Emergency Rooms in England) activity data and admitted patient care.
In 2018 and 2020, the US National Institute of Standards and Technology (NIST) ran two challenges related to synthetic data: the Differential Privacy \href{https://www.nist.gov/ctl/pscr/open-innovation-prize-challenges/past-prize-challenges/2018-differential-privacy-synthetic}{Synthetic Data} and \href{https://www.nist.gov/ctl/pscr/open-innovation-prize-challenges/past-prize-challenges/2020-differential-privacy-temporal}{Temporal Map} challenges, awarding cash prizes seeking innovative synthetic data algorithms and metrics.

\section*{Risks of Using Synthetic Data}
To reason around the risks of synthetic data, researchers have used a few ``metrics'' to measure privacy properties.

\descr{Linkage.}
Because synthetic data is ``artificial,'' a common argument is that there is no direct link between real and synthetic records, unlike anonymized records.
Thus, researchers have used similarity tests between real and synthetic records to support the safety of synthetic data.
Unfortunately, however, this kind of metric fails to grasp the real risks of a strategic adversary using features that are likely to be influenced by the target's presence. %

\descr{Attribute Disclosure.}
This kind of privacy violation happens whenever access to data allows an attacker to learn {\em new} information about a specific individual~\cite{hittmeir2020baseline}, e.g., the value of a particular attribute like race, age, income, etc.
Unfortunately, if the real data contains strong correlations between attributes, these correlations will likely be replicated in the synthetic data and available to the adversary~\cite{stadler2022synthetic}.
Furthermore, Theresa Stadler et al.~\cite{stadler2022synthetic} show that records with rare attributes or whose presence affects the ranges of numerical attributes remain highly vulnerable to disclosure.

\descr{From metrics to attacks.}
Roughly speaking, linkage is often formulated as a successful {\em membership inference} attack, whereby an adversary aims to infer if the data from specific target individuals were relied upon by the synthetic data generation process.

\begin{figure}[h]
\centering
\includegraphics[width=0.8\columnwidth]{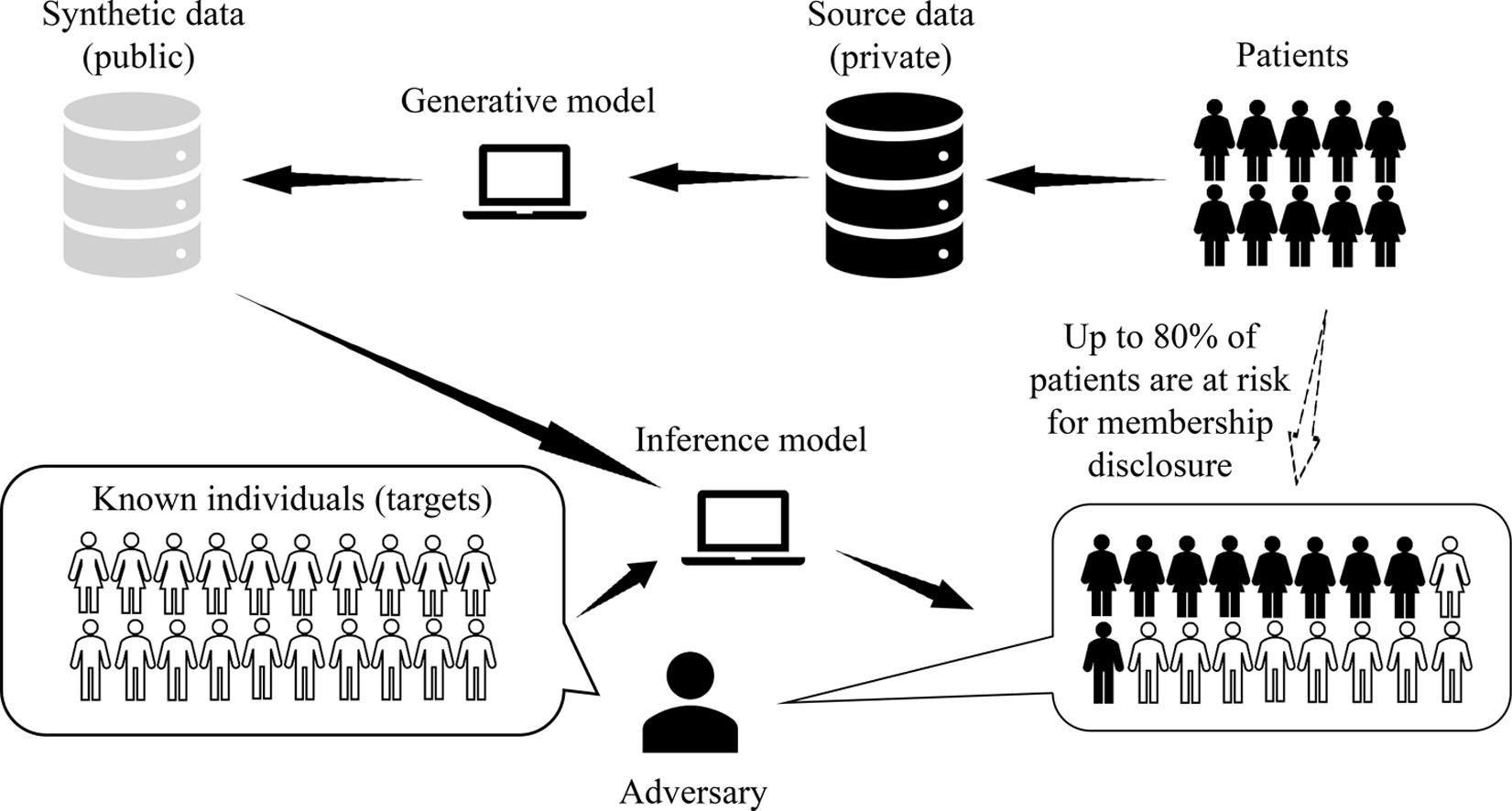}
\caption{Membership Inference Attack (Source: \cite{brad})}
\label{fig:attack}
\end{figure}

\noindent Consider the example in Figure~\ref{fig:attack},  where synthetic health images are used for research: discovering that a specific record was used in a study leaks information about the individual's health.

Attribute disclosure is usually formulated as an {\em attribute/property inference} attack.
Here, the adversary, given some public information, tries to reconstruct some private attributes of some target users.

\begin{figure}[h]
\centering
\includegraphics[width=0.9\columnwidth]{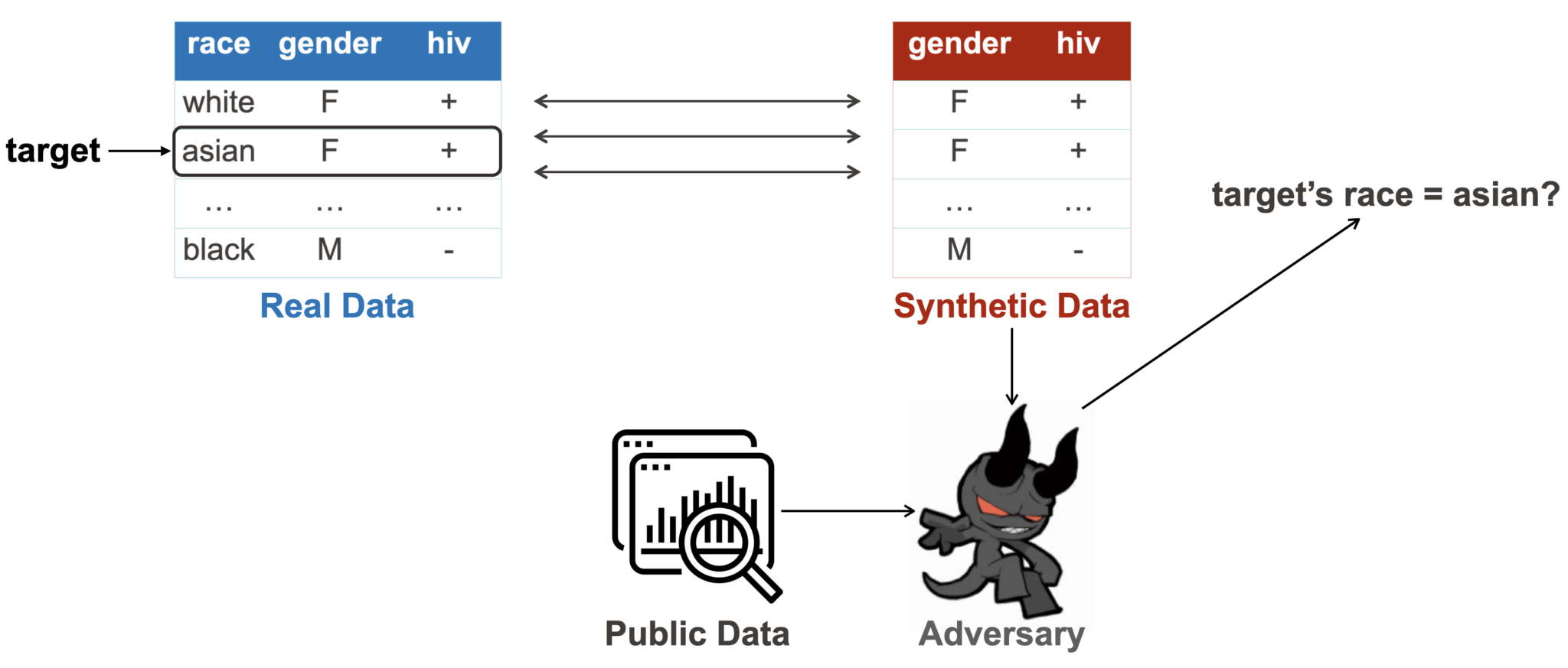}
\caption{Attribute Inference Attack.}
\end{figure}

\descr{How realistic are the attacks?}
One important consideration that applies to most privacy studies is that they do not provide {\em ``binary''} answers, e.g., making definite statements of whether some method provides perfect privacy or none at all.
Rather, they provide probability distributions vis-\`a-vis different systems/threat models, adversarial assumptions, datasets, etc.

However, there still are a significant number of gaps identified by state-of-the-art research studies.
In practice, synthetic data provides little additional protection compared to anonymization techniques, with privacy-utility trade-offs being even harder to predict~\cite{stadler2022synthetic}.

\descr{Differentially Private Synthetic Data Generation.}
The state-of-the-art method for providing access to information free from inferences is to satisfy %
differential privacy~\cite{nissim2017differential}.

In the context of synthetic data, the intuition is to train the generative models used to produce synthetic data in a differentially private manner.
Typically, one of three methods is used: using the Laplace mechanism, sanitizing the gradients during stochastic gradient descent, or using a technique called PATE, which transfers to a ``student'' model the knowledge of an ensemble of ``teacher'' models (privacy is provided by training teachers on disjoint data and noisy aggregation of teachers' answers)~\cite{PATE}. 
The resulting methods tend to combine generative model architectures with differential privacy; state-of-the-art tools include DP-GAN~\cite{xie2018differentially}, DP-WGAN~\cite{dpwgan}, DP-Syn~\cite{li2021dpsyn}, PrivBayes~\cite{privBayes}, PATE-GAN~\cite{jordon2018pate}, etc. 
A list of relevant papers (with code) is available on Georgi Ganev's \href{https://github.com/ganevgv/dp-generative-models}{GitHub page}.

\section*{The Inherent Security and Privacy Limitations}
While there likely are other challenges related to synthetic data (e.g., regarding usability, fidelity, and interpretability), we focus on security and privacy once.
In particular, with respect to privacy, we believe it unlikely that synthetic data will provide a silver bullet to sanitize sensitive data or safely share confidential information across the board.
Instead, there could be specific use cases where training a generative model provides better flexibility and privacy protection than the alternatives.
For instance, financial companies can use synthetic data to ensure production data is not used during testing or shared across different sub-organizations.
Also, governmental agencies could enable citizens and entities to extract high-level statistics from certain data distributions. %

However, we argue that those case studies are not going to generalize.
Essentially, generative models trained without differential privacy (or with very large privacy budgets) do not provide high safety, privacy, or confidentiality levels.
Conversely, differentially private synthetic data generation algorithms can but with a non-negligible cost to utility/accuracy.
More precisely, protecting privacy inherently means you must ``hide'' vulnerable data points like outliers, etc.
So if one tries to use synthetic data to upsample an under-represented class, train a fraud/anomaly detection model, etc., they will inherently need to choose between {\em either} privacy {\em or} utility.

Another limitation is that usable privacy mechanisms must be %
{\em predictable}, i.e., %
enabling reliable assumptions of how parties will access and use personally identifiable information~\cite{nist}. 
That is not always the case with synthetic data, because of the probabilistic nature of generative models and the inherent difficulty of predicting what signals a synthetic dataset will preserve and what information will be lost~\cite{stadler2022synthetic}.

\section*{Looking Ahead}
In conclusion, we believe there are several interesting research questions that the community should be focusing on in the context of synthetic data:

\begin{enumerate}
\item While differential privacy often provides an overly conservative, worst-case approach to privacy, it allows abstracting away from any adversarial assumption.
But in practice, the accuracy of the attacks we can realistically mount is measurably far from the theoretical bounds, which motivates the need for more work on auditing differentially private synthetic data generation algorithms.

\item We call for the privacy engineering community to help practitioners and stakeholders identify the use cases where synthetic data can be used safely, perhaps even in a semi-automated way.

\item Researchers should also be incentivized to provide actionable guidelines to understand the distributions, types of data, tasks, and settings, where one could achieve reasonable privacy-utility tradeoffs via synthetic data. %
\end{enumerate}

\subsection*{Acknowledgements}
\noindent We wish to thank Georgi Ganev, Bristena Oprisanu, and Meenatchi Sundaram Muthu Selva Annamalai for reviewing a draft of this article.

{\small
\bibliographystyle{unsrt}

}
\end{document}